\documentclass[reprint, aps]{revtex4-1}
\usepackage[version=3]{mhchem} % Formula subscripts using \ce{}

\usepackage{amsmath}  
\usepackage{amsfonts} 
\usepackage{subfig}
\usepackage{graphicx} 
\usepackage{booktabs}
\usepackage{threeparttable}
\usepackage[font=footnotesize]{caption}
\usepackage{verbatim}

\newcommand{\mbf}[1]{\mathbf{#1}}

\begin{document}

%% Each author should be given as a separate \author command.
%%
%% Corresponding authors should have an e-mail given after the author
%% name as an \email command. Phone and fax numbers can be given
%% using \phone and \fax, respectively; this information is optional.
%%
%% The affiliation of authors is given after the authors; each
%% \affiliation command applies to all preceding authors not already
%% assigned an affiliation.
%%
%% The affiliation takes an option argument for the short name.  This
%% will typically be something like "University of Somewhere".
%%
%% The \altaffiliation macro should be used for new address, etc.
%% On the other hand, \alsoaffiliation is used on a per author basis
%% when authors are associated with multiple institutions.
%%%%%%%%%%%%%%%%%%%%%%%%%%%%%%%%%%%%%%%%%%%%%%%%%%%%%%%%%%%%%%%%%%%%%
\author{James Shee}
\email{js4564@columbia.edu}
\affiliation{Department of Chemistry, Columbia University, 3000 Broadway, New York, NY, 10027}
\author{Evan J. Arthur}
\affiliation{Schr\"odinger Inc., 120 West 45th Street, New York, New York 10036, United States}
\author{Shiwei Zhang}
\affiliation{Department of Physics, College of William and Mary, Williamsburg, Virginia 23187-8795}
\author{David R. Reichman}
\author{Richard A. Friesner}
\affiliation{Department of Chemistry, Columbia University, 3000 Broadway, New York, NY, 10027}

%%%%%%%%%%%%%%%%%%%%%%%%%%%%%%%%%%%%%%%%%%%%%%%%%%%%%%%%%%%%%%%%%%%%%
%% The document title should be given as usual. Some journals require
%% a running title from the author: this should be supplied as an
%% optional argument to \title.
%%%%%%%%%%%%%%%%%%%%%%%%%%%%%%%%%%%%%%%%%%%%%%%%%%%%%%%%%%%%%%%%%%%%%
\title{Phaseless Auxiliary-Field Quantum Monte Carlo on Graphical Processing Units}

\begin{abstract}
We present an implementation of phaseless Auxiliary-Field Quantum Monte Carlo (ph-AFQMC) utilizing graphical processing units (GPUs).  The AFQMC method is recast in terms of matrix operations which are spread across thousands of processing cores and are executed in batches using custom Compute Unified Device Architecture kernels and the hardware-optimized cuBLAS matrix library.  Algorithmic advances include a batched Sherman-Morrison-Woodbury algorithm to quickly update matrix determinants and inverses, density-fitting of the two-electron integrals, an energy algorithm involving a high-dimensional precomputed tensor, and the use of single-precision floating point arithmetic.  These strategies result in dramatic reductions in wall-times for both single- and multi-determinant trial wavefunctions.  For typical calculations we find speed-ups of roughly two orders of magnitude using just a single GPU card.  Furthermore, we achieve near-unity parallel efficiency using 8 GPU cards on a single node, and can reach moderate system sizes via a local memory-slicing approach.  We illustrate the robustness of our implementation on hydrogen chains of increasing length, and through the calculation of all-electron ionization potentials of the first-row transition metal atoms.  We compare long imaginary-time calculations utilizing a population control algorithm with our previously published correlated sampling approach, and show that the latter improves not only the efficiency but also the accuracy of the computed ionization potentials.  Taken together, the GPU implementation combined with correlated sampling provides a compelling computational method that will broaden the application of ph-AFQMC to the description of realistic correlated electronic systems. 
\end{abstract}

% for Arxiv format
\maketitle


%%%%%%%%%%%%%%%%%%%%%%
\section{Introduction}
%\section{I. Introduction}
%%%%%%%%%%%%%%%%%%%%%%

Auxiliary-Field Quantum Monte Carlo (AFQMC) is a computational method capable of predicting ground-state observables of chemical systems with very high accuracy.\cite{motta2017ab,zhang1995constrained,zhang2003quantum,al2006auxiliary,al2006oxides,al2006postd,al2007bond,suewattana2007phaseless,al2007Hbonds,purwanto2008eliminating,purwanto2009pressure,purwanto2011assessing,virgus2012ab,purwanto2013frozen,virgus2014stability,purwanto2015auxiliary,ma2015quantum,purwanto2016auxiliary,ma2017auxiliary,shee2017chemical,motta2017Hchain,motta2017computation,zheng2017stripe}  For finite-sized systems such as molecules in Gaussian basis sets, AFQMC calculations scale with the fourth power of the system size, which compares favorably with traditional wavefunction methods such as second-order M\o ller-Plesset Perturbation Theory (MP2),\cite{moller1934note} Coupled Cluster (CC) approaches,\cite{bartlett2007coupled} and Complete Active Space methods such as CASSCF\cite{roos2007complete} and CASPT2.\cite{andersson1992second}  However, the prefactor of a typical AFQMC calculation is relatively large.  Recently we have introduced a correlated sampling (CS) approach for quantities involving energy differences which is capable of reducing computational prefactors by approximately an order of magnitude.\cite{shee2017chemical}  In this work we present a different but complementary strategy involving hardware optimization on graphical processing units (GPUs) which can drastically reduce the prefactors in calculations of general ground-state properties.  

GPUs have several distinct advantages over traditional Central Processing Units (CPUs), including the ability to perform hardware-optimized matrix operations in serial and in ``batches," and the use of single-precision (sp) floating-point arithmetic with significant gains in computational speed.  In recent years the use of GPUs has been extended well beyond traditional image visualization tasks into many fields such as machine learning\cite{lecun2015deep} and molecular mechanics.\cite{anderson2008general}  Of particular relevance to our work presented here is the progress in performing electronic structure calculations on GPUs. This hardware has been utilized to efficiently evaluate the integrals required in ab-initio calculations,\cite{ufimtsev2008quantum,luehr2011dynamic,kalinowski2017arbitrary,kussmann2017hybrid}
\begin{comment}
Integral evaluation
	GPU evaluation of 2e integrals\cite{ufimtsev2008quantum} 
    hybrid CPU and GPU integral generation engine,\cite{kussmann2017hybrid} 
    2e integrals with arbitrary angular momentum (up to f on GPU).  Factors up to 30x vs a single serial CPU found for hybrid GPU CPU HF method, with up to 3k basis functions. illustrate on alanine chains, taxol, vancomycin, C180, vancomycin dimer, olestra.\cite{kalinowski2017arbitrary} 
    Dynamic precision GPU evalualation of 2e integrals\cite{luehr2011dynamic} 
\end{comment}
to perform Hartree-Fock\cite{ufimtsev2008graphical,yoshikawa2015linear} (HF) and
\begin{comment}
	HF calculations with divide and conquer local (linear scaling) methods with RI techniques to overcome memory bottlenecks.  3-10x speedups. Can handle small proteins in triple zeta basis.\cite{yoshikawa2015linear} 
    2008 review.  Table 3 shows 32 bit (sp) arithmetic produces scf energies within chemical accuracy for things like cafffeine and cholesterol (26x and 55x speedup respectively).  However, larger molecules (buckyball, taxol, valinomycin) were not within chemical accuracy, though they displayed larger speedups.\cite{ufimtsev2008graphical}
\end{comment}
Density Functional Theory (DFT) calculations,\cite{yasuda2008accelerating,genovese2009density,hacene2012accelerating}
\begin{comment}
	large molecules - taxol and valinomycin - GPU algorithm for DFT with mostly single precision about a factor of 10 faster, with negligible error from sp parts.\cite{yasuda2008accelerating}
    DFT on hybrid GPU/CPU\cite{genovese2009density} 
    real-space DFT (as opposed to GTO-based), comparable run-time to GTO implementation in Terachem.\cite{andrade2013real} 
    VASP plane wave DFT.  factors of between 3 and 8 on n (CPUcore+GPU) vs n CPUs alone.\cite{hacene2012accelerating}
\end{comment}
and to study model systems such as the Hubbard Model within the dynamical cluster approximation\cite{meredith2009accuracy} and the Ising model.\cite{levy2010simulating,block2010multi}
\begin{comment}
	HTSC study, 2D Hubbard Model with DCA (impurity problem, QMC as solver).  Computed critical temperature, comparing GPU sp and CPU sp and dp.  Within 0.15 percent.  sp had slightly larger variance.  Transfer minimization involved "Matrix Handler class".  Total speedup of 5x\cite{meredith2009accuracy} 
    lattice spin models.  2D Ising model .  70-150x speedups over single-threaded codes.\cite{levy2010simulating} 
    2d Ising model, 35x compared to 1 CPU.\cite{block2010multi} 
\end{comment}
In addition there have been recent GPU implementations of MP2,\cite{olivares2009accelerating,doran2016monte,song2016atomic,vogt2008accelerating}
\begin{comment}
	Table 1 shows less than 1 kcal/mol error from single-precision for systems up to C32H66, RI-MP2.  Propose a mixed precision matrix multiplication.\cite{olivares2009accelerating}
    MC-MP2 on 256 GPUs\cite{doran2016monte} 
    Tensor hyper-contracted and SOS (Scaled Opposite Spin) MP2 on GPUs, with analytical gradients.  Has a lot of good references to DF techniques.\cite{song2016atomic}  
    RI-MP2.  alkane chains with even number of C atoms (from C8 to C22), cc-pVDZ basis set.  for latter, 4.5x speedup was found.   single precision errors RMSD=0.5mHa, MAD=0.3mHa.  this is encouraging\cite{vogt2008accelerating} 
\end{comment}
CC methods,\cite{deprince2011coupled} 
\begin{comment}
	spin-free CCSD on GPUs. 4-5x speedup relative to multithreaded CPU algorithm on same-generation hardware.  Single-precision errors on order of 1e-6 Hartrees, 1e-5 at worst.  Considered polyacetylene series (CnHnplus2), acene series for 2,3,4 fused benzene rings, C20 fullerene with 6-31G basis.  For large systems (e.g. C18H20) use tiling (subset of integrals, as much as 3GB, pushed to GPU every iteration) and still found 4x speedup, and for more accuracy use a mixed precision strategy (diff precision for diff iterations).  They have memory limit of 200 spatial orbitals.  ``Regardless of these arguments for tiled matrix multiplications, the memory limitation is completely artificial in the sense that it is coupled to current hardware limitations and will therefore change as GPU hardware matures for scientific applications; the NVIDIA C2070 card, which was not used in these experiments, has 6 GB of global memory."\cite{deprince2011coupled}
\end{comment}
TDDFT,\cite{isborn2011excited} Configuration Interaction (CI),\cite{fales2015nanoscale,fales2017complete} and CASSCF approaches.\cite{hohenstein2015atomic,snyder2017direct}
\begin{comment}
	Configuration Interaction with Singles and TDDFT\cite{isborn2011excited} 
	FCI/CASCI GPU implementation.\cite{fales2015nanoscale,fales2017complete} 
	CASSCF GPU implementation for large system sizes and active spaces.  1000 basis functions, and apparent quadratic scaling.\cite{hohenstein2015atomic,snyder2017direct} 
\end{comment}
Efficient algorithms to compute energy gradients\cite{ufimtsev2009quantum,song2017analytical} and tensor contractions\cite{kaliman2017new} have also been developed.
\begin{comment}
Tensor contractions using GPUs.  Library integrated with Q-Chem.  Enable canonical CC and EOM CCSD with over 1000 basis functions.\cite{kaliman2017new} 
\end{comment}

With respect to Quantum Monte Carlo (QMC) methods, GPU implementations have been formulated primarily for real-space approaches.  For example, Diffusion Monte Carlo (DMC) with sp arithmetic has been accelerated by a factor of $\sim$6x on a GPU versus a quad-core CPU.\cite{anderson2007quantum}  A recent study employing a multi-GPU implementation has reported speed-ups of a factor of 10-15x relative to a quad-core CPU for Variational and Diffusion MC for real materials.\cite{esler2012accelerating} Very recently an open-source QMC suite, QMCPACK,\cite{kim2018qmcpack} has released scalable implementations of real-space methods including Variational, Diffusion, and Reptation MC.  An implementation of auxiliary-field QMC is mentioned in Ref. \citenum{kim2018qmcpack}, although data illustrating its efficiency and accuracy is not yet available.

In this paper we detail our GPU implementation of the phaseless variant of auxiliary-field QMC (ph-AFQMC), and illustrate its performance and accuracy via calculations of the total energies of linear chains of hydrogen atoms and the all-electron ionization potentials (IPs) of the first-row transition metal (TM) atoms.  We explicitly compare our GPU wall-times with comparable CPU timings. Speed-ups from the GPU port of two orders of magnitude are seen in large systems, with the potential for even greater reductions of the scaling prefactor depending on the system-size.  The robustness and accuracy of our implementation are shown by comparing our calculated values to either exact numerical techniques or experiment.

Our paper is organized as follows: In Sec. II we provide a concise review of AFQMC and the phaseless constraint.  In Sec. III we detail our GPU implementation and highlight significant algorithmic additions.  In Sec. IV and V we present timing and accuracy results for the hydrogen chains and TM IPs, respectively, and comment on the advantages of the correlated sampling approach.  In Sec. VI we conclude with a summary of our results and a discussion of future work.

%%%%%%%%%%%%%%%%%%%%%%%%%%%%%%%%%
\section{Brief Overview of AFQMC}
%\section{II.  Brief Overview of AFQMC}
%%%%%%%%%%%%%%%%%%%%%%%%%%%%%%%%%

Detailed expositions of AFQMC and the phaseless constraint can be found elsewhere.\cite{zhang2003quantum, motta2017ab, shee2017chemical}  In this section we  highlight only aspects that are directly relevant to this work.

The ground state of a many-body wavefunction $| \Phi \rangle$ can be obtained via imaginary-time propagation, \emph{i.e.}  $\lim_{N\rightarrow\infty}(e^{\Delta\tau(E_0 - \hat{H})})^N |\Phi\rangle = |\Phi_0 \rangle$, if $\langle \Phi_0 | \Phi \rangle \ne 0$.  The general electronic Hamiltonian is
\begin{equation}
\hat{H} = \sum\limits_{ij}^M \ T_{ij} \sum\limits_\sigma c_{i \sigma}^\dagger c_{j \sigma} + \frac{1}{2}\sum\limits_{ijkl}^M V_{ijkl} \sum\limits_{\sigma,\tau} c_{i \sigma}^\dagger c_{j \tau}^\dagger c_{l\tau} c_{k\sigma},
\label{Hamiltonian}
\end{equation}
where $M$ is the size of an orthonormal one-particle basis, $\sigma$ and $\tau$ denote electron spin, and $c_{i\sigma}^\dag$ and $c_{i\sigma}$ are the second-quantized fermionic creation and annihilation operators.  The matrix elements, $V_{ijkl}$, can be expressed as a sum over products of three-index quantities via Cholesky decomposition\cite{purwanto2011assessing} (CD) or Density-Fitting\cite{whitten1973coulombic} (DF) procedures, allowing  Eq. \eqref{Hamiltonian} to be written as the sum of all one-body operators plus a two-body operator of the form $\hat{H}_2 = -\frac{1}{2}\sum_\alpha \hat{v}_\alpha^2$.  

After utilizing a Trotter-Suzuki decomposition\cite{trotter1959product} and the Hubbard-Stratonovich transformation written in the form\cite{stratonovich1957method, hubbard1959calculation}
\begin{equation}
e^{\frac{1}{2}\Delta\tau \hat{v}_\alpha^2} = \int_{-\infty}^\infty dx_\alpha \bigg(\frac{e^{-\frac{1}{2}x_\alpha^2}}{\sqrt{2\pi}}\bigg)e^{\sqrt{\Delta\tau}x_\alpha \hat{v}_\alpha}, 
\label{HS}
\end{equation}
the imaginary-time propagator can be expressed as an exponential of a one-body operator integrated over auxiliary-fields
\begin{equation}
e^{-\Delta\tau \hat{H}} = \int d\mbf{x} P(\mbf{x}) \hat{B}(\mbf{x}).  
\label{prop}
\end{equation}
This integral is evaluated using a Monte Carlo technique in which the wavefunction is represented as a weighted sum of Slater determinants.  A theorem due to Thouless\cite{thouless1960stability} enables the propagation of the Slater determinants along ``paths" defined by a set of auxiliary-fields, which in practice can be accomplished via matrix-matrix multiplications.  Each auxiliary-field is shifted by a force-bias chosen to minimize the fluctuations in a walker's weight.  With this choice the weights are updated after each propagation step by a factor proportional to $e^{-\Delta\tau E_L}$, where the local energy is defined as $E_L = \frac{\langle \psi_T|\hat{H}|\phi\rangle}{\langle \psi_T | \phi \rangle}$, and $| \psi_T \rangle$ is a trial wavefunction, typically taken as a single Slater determinant or a linear combination of determinants.

For the Coulomb interaction, the propagator in Eq. \eqref{prop} has an imaginary component which rotates the walker orbitals and weights in the complex plane, leading inevitably to divergences and an exponentially decaying signal to noise ratio.  The phaseless approximation can be used to constrain the weights to be real and positive.  This involves taking the real-part of the local energy, and multiplying the resulting weights by $max\{0, cos(\Delta\theta) \}$, where $\Delta\theta = Im \{ln \frac{\langle \psi_T | \phi^{\tau + \Delta\tau} \rangle}{\langle \psi_T | \phi^\tau \rangle} \}$.  The latter breaks the rotational symmetry of the random walk in the complex plane by choosing a unique gauge for $|\Phi_0\rangle$ (and eliminates walkers whose weights have undergone phase rotations in excess of $\pm \frac{\pi}{2}$).  We note that the severity of the phaseless constraint can be systematically reduced with the use of trial wavefunctions which even more closely represent the true ground-state.

%%%%%%%%%%%%%%%%%%%%%%%%%%%%
\section{GPU Implementation}
%\section{III. GPU Implementation}
%%%%%%%%%%%%%%%%%%%%%%%%%%%%

In contrast to traditional computing paradigms which utilize CPUs to execute all computing tasks, we employ a strategy in which CPUs offload a majority of the computational effort to one or more GPU cards.  A typical GPU device has 4-12 GB of memory which is separate from that accessible by the CPU; therefore, data must be allocated on both the host and device, and inter-communication between these different memory spaces requires the explicit copying of data back and forth.  In this work great care is taken to minimize such transfers, and we create custom memory structures that organize memory addresses and facilitate switching between sp and double-precision (dp).

Once the relevant data resides on the GPU(s), we simply call Nvidia's Compute Unified Device Architecture (CUDA) Basic Linear Algebra Subprograms (cuBLAS) library whenever possible, e.g., to execute the matrix-matrix multiplications that propagate walker determinants.  However, our largest speed-ups relative to CPU execution are obtained by writing custom C/CUDA functions for all significant parts of the code, e.g. calculation of the force-bias, local energy, etc.   

\begin{figure}[h!]
    \centering
    \includegraphics[width=9cm]{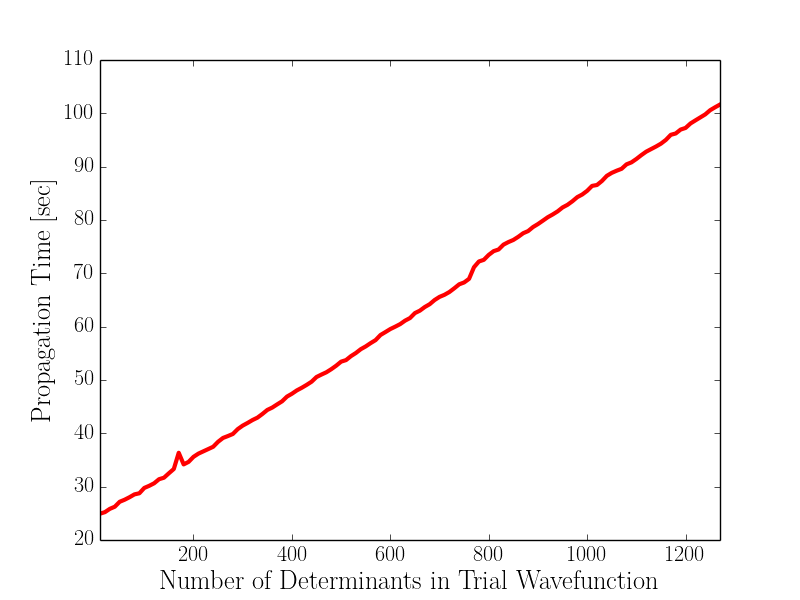} 
    \caption{Propagation time vs the number of configurations in the CASSCF trial wavefunction for the Mn atom in the aug-cc-pwCVQZ-DK basis.  Calculations use sp, a CD threshold of $10^{-4} Ha$, 20 walkers and imaginary-time trajectories of length 1 $Ha^{-1}$ with a time step of $\Delta\tau = 0.005 \ Ha^{-1}$. Walker orthonormalization and local energy measurements were performed every 2 and 20 steps, respectively.}   
    \label{fig:Mn_Ndet}
\end{figure}

A relatively new addition to the cuBLAS library are so-called "batched" functions which perform many smaller operations simultaneously, e.g. a set of matrix-matrix multiplies or lower-upper (LU) decompositions. These batched functions are well-suited for operations that are individually too small to parallelize effectively across thousands of cores.  We utilize this feature heavily in our implementation of Sherman-Morrison-Woodbury (SMW) updates to quickly compute equal-time Green's functions when multi-determinant trial functions are used.  We note that previous Diffusion MC studies have utilized similar SMW updates,\cite{clark2011computing,mcdaniel2017delayed,esler2012accelerating} and Ref. \citenum{shi2018accelerating} has also presented fast updates and memory-saving techniques for multi-determinant CI trial wavefunctions in AFQMC.  Given a reference matrix $A$, the following formulas are used to compute the determinants and inverses of matrices which differ from $A$ by one or more row or column:
\begin{equation}
\begin{split}
det(A + U_i V_i^T) & = det(I + V_i^T A^{-1}U_i) det(A) \\ 
(A + U_i V_i^T)^{-1} & = A^{-1} - A^{-1}U_i(I + V_i^T A^{-1} U_i)^{-1} V_i^T A^{-1}.
\end{split}
\end{equation}
In the context of ph-AFQMC, suppose we use a multi-determinant trial wavefunction, $|\Psi_T\rangle = \sum_{i=0} c_i |\psi_{T,i} \rangle$, where $\langle \psi_{T,i} | \psi_{T,j} \rangle = \delta_{ij}$.  Then, for the $k^{th}$ walker determinant $|\phi_k\rangle$, $A = [ \psi_{T,i=0}]^\dag [\phi_k ]$, where the square brackets denote a matrix representation, $U_i, V_i$ are of dimension $N_\sigma$ x $E_i$, where $N_\sigma$ is either the number of spin-up or spin-down electrons, and $E_i$ is the number of excitations required to form the $i^{th}$ configuration of the multi-determinant expansion from the reference configuration.  The determinant and inverse of the reference matrix corresponding to zero excitations ($i = 0$) is computed first for spin-up and spin-down configurations, followed by batched SMW updates for all $i \ne 0$.  Sub-cubic scaling with respect to particle number is achieved since $E_i << N_\sigma$.  Fig. \ref{fig:Mn_Ndet} highlights the efficiency of our batched implementation of the SMW algorithm, for the Mn atom in the aug-cc-pwCVQZ-DK basis (185 basis functions, 25 electrons).  ``Propagation time" denotes the total wall-time minus the time spent on initial setup, e.g., memory allocation, input/output, and precomputation of the operators and required intermediates.  Previously, going from, e.g., 10 to 1200 determinants would multiply the propagation time by a factor of 120.  In contrast, our SMW algorithm reduces this to a mere factor of 3.9. 

We have developed a GPU-optimized algorithm for evaluating the local energy of a walker.  4-index tensors are precomputed once at the start of a simulation, which, in the spin-free case (for simplicity), are of the form: 
\begin{equation}
Y_{ijab} = \sum_{kl} \sum_\alpha L_{ik}^\alpha L_{jl}^\alpha [\psi_{T, ak}^{\dag} \psi_{T, bl}^{\dag} - \psi_{T, al}^{\dag} \psi_{T, bk}^{\dag}],
\label{Ys}
\end{equation}
where the $L^\alpha$ arise from decomposing the two-electron integrals via CD or DF ($V_{ijkl} = \sum_\alpha L_{ik}^\alpha L_{jl}^\alpha$), indices $i,j,k,l$ run from 1 to $M$ (basis size) while indices $a,b$ run from 1 to $N$ (number of electrons).  $\psi_{T}$ is a matrix with columns composed of the orbitals in the trial function.  Importantly, the sums over $k,l$ and over auxiliary-fields $\alpha$, the number of which typically scales as 2-10$M$, need only be computed once at the start of the simulation.  Using Eq. \eqref{Ys} the energy expression is then recast in terms of large matrix operations in order to realize maximal GPU-acceleration.  

Finally we introduce the use of DF in AFQMC calculations, where effective densities $\bar{\rho}_{ij}(\mbf{r})$ are fit to auxiliary basis functions, $\chi(\mbf{r})$: 
\begin{align}
V_{ijkl} & = \int d\mbf{r}_1 d\mbf{r}_2 \phi_i(\mbf{r}_1 ) \phi_j(\mbf{r}_1) \frac{1}{r_{12}} \phi_k(\mbf{r}_2 )\phi_l(\mbf{r}_2) \\
            & \sim \int d\mbf{r}_1 d\mbf{r}_2 \bar{\rho}_{ij}(\mbf{r}_1) \frac{1}{r_{12}} \bar{\rho}_{kl}(\mbf{r}_2),
\end{align}
where $\bar{\rho}_{ij} (\mbf{r}) = \sum_\mu d_\mu^{ij} \chi_\mu (\mbf{r})$.  By construction, $\mu < M$, typically reducing the number of auxiliary-fields in AFQMC to $\sim$2$M$.  In addition to reducing the memory footprint and speeding up the calculation of the force-bias and the assembly of the one-body operator in Eq. \eqref{HS}, fewer auxiliary-fields generally leads to a reduction in statistical noise.  

%--------------------------------------------%
\section{Illustration with Hydrogen Chains}
%\section{IV. Illustration with Hydrogen Chains}
%--------------------------------------------%

In this section we explore the effects on both computational efficiency and accuracy due to the use of sp vs dp, and DF vs CD for linear chains of hydrogen atoms.  These systems have played an important role in benchmarking new theories of correlated electronic materials.\cite{hachmann2006multireference,tsuchimochi2009strong,al2007bond,sinitskiy2010strong,lin2011dynamical,stella2011strong,motta2017Hchain}  While these systems do not capture many nuances of more realistic molecular systems, they are nevertheless a useful prototype capable of (1) yielding wall-time and scaling insights due to the ability to systematically increase the system size, (2) providing an atomistic analogue of well-studied model systems such as the Heisenberg and Hubbard models albeit with a more realistic description of  long-range Coulomb interactions, while (3) exhibiting strong static correlation at large bond lengths. 

\subsection{Computational Details}
%\subsection{IV.A  Computational Details}
For all hydrogen chain calculations we use the cc-pVDZ basis, for which there is abundant benchmark data.\cite{motta2017Hchain}  In this basis there are 5 basis functions per electron, a notably smaller number than used in typical molecular calculations. The Weigend Coulomb-fitting basis set\cite{weigend2006accurate} is employed as the auxiliary basis for DF, and CDs in this section employ a threshold of $10^{-5}Ha$ (as chosen in Ref. \citenum{motta2017Hchain}).  

We use PySCF\cite{sun2018pyscf} to compute all inputs required of our ph-AFQMC code.  Unless otherwise specified we use an imaginary-time step of 0.005 $Ha^{-1}$.  Walker orbitals are orthonormalized after every two propagation steps, to preserve the anti-symmetry of the walker configurations and also to keep the magnitude of orbital coefficients and associated quantities as small as possible (thus extending the accuracy of sp).  We employ the hybrid method of ph-AFQMC\cite{purwanto2009pressure} to minimize evaluations of the local energy, which is measured every 0.1 $Ha^{-1}$.  The total number of walkers is fixed throughout each simulation, and when required we use a population control (PC) algorithm at intervals of 0.1 $Ha^{-1}$.  Long imaginary-time runs utilizing PC use a reblocking analysis\cite{reblocking} to obtain statistical errors uncontaminated by autocorrelation.  All calculations are run on Nvidia GeForce GTX 1080 GPUs, with 2.1 GHz Intel Xeon CPUs. 

%------------------------%
\subsection{Timings}
%\subsection{IV.B Timings}
%------------------------%
   
\begin{figure}[h!]
\centering
\includegraphics[width=0.5\textwidth]{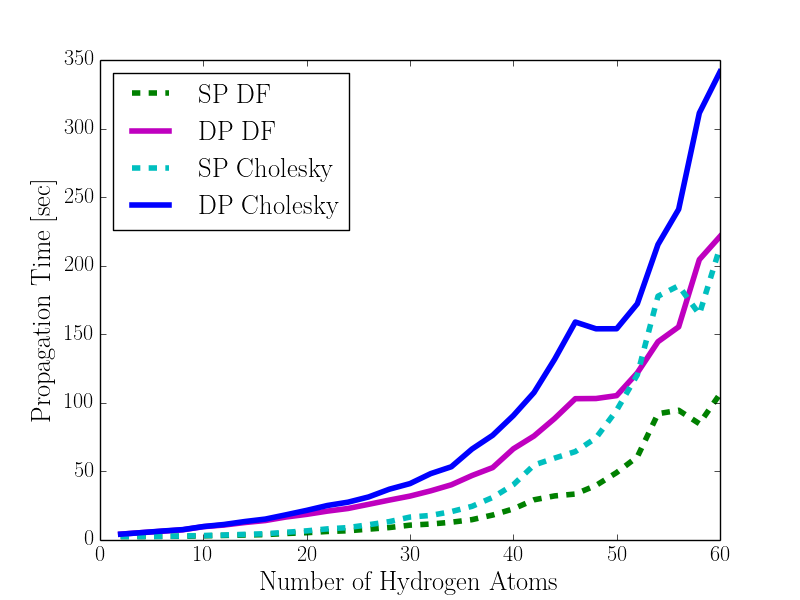}
\caption{Propagation time using 1 GPU for hydrogen chains of varying lengths, comparing two types of two-electron integral decompositions, DF vs CD with a $10^{-5}$ cutoff, within both sp and dp.  UHF trial functions are used, and 24 walkers are propagated for an imaginary-time segment of length 1 $Ha^{-1}$.}
\label{fig:GPUflavours}
\end{figure}

Employing an unrestricted HF trial for ph-AFQMC has been shown to produce very accurate energies for hydrogen chains near their equilibrium bond lengths.\cite{motta2017Hchain}  Using a bond length of 1.880(2) Bohr as given by Density Matrix Renormalization Group (DMRG) in the cc-pVDZ basis, we compare propagation times using a single GPU card for an increasing number of hydrogen atoms.  Sample propagation times for several variants of precision and means of decomposing the two-electron terms are shown in Fig. \ref{fig:GPUflavours}.  For $H_{60}$ DF is 2.0x faster than CD in sp and 1.5x faster in dp.  Sp is 2.1x faster than dp when DF is used, and 1.6x faster using CD.  Generally we find that the relative speed-ups afforded by sp over dp, and DF over CD, increase with system size.  The non-monotonicity of the propagation times vs system size is a unique and rather unexpected artifact of the GPU architecture, and we observe that the sp (dotted) and dp (lines) trajectories move together, suggesting a different treatment of sp and dp at the hardware level.  At this time our CPU code optimized with the SMW formula and precomputing algorithms is not able to treat unrestricted trial functions.  For the sake of fair comparisons we will compare CPU vs GPU timings only for restricted (multi-determinant) trials.

In Table \ref{table:Ndet_timings_Hchain} we benchmark the performance of ph-AFQMC with CD using multi-determinant trial functions.  Compared to our CPU code in dp, which also uses SMW and precomputing in the measurements of the local energies, our GPU-accelerated code in sp achieves large speed-ups ranging from 87.2x with two determinants to 670.1x with 1000 determinants.  Importantly we find that the relative speed-up increases with the number of determinants present in the trial function.  Apart from roughly a factor of 2 in time due to the use of sp on the GPU, the chief source of the speed-ups is the batched nature of the evaluation of mixed-expectation values involving the trial wavefunction.

\begin{table*}[ht]
%\begin{table*}[h!]
%\centering
\begin{threeparttable}
\caption{Propagation times (in seconds) for an H$_{50}$ chain with a varying number of determinants that comprise the trial wavefunction.  We use CD with a $10^{-5}$ cutoff, and show the speed-up of a single GPU in sp over a single CPU in dp.}
\begin{tabular}{l l l l l l}
\hline\hline
         & $N_{det} = 2$ & $N_{det} = 50$ & $N_{det} = 100$ &  $N_{det} = 500$ & $N_{det} = 1000$  \\ [0.5ex] 
\hline
GPU sp   & 99.9   & 105.3   & 111.8   &  158.0  &  218.7    \\ 
CPU dp   & 8775.2 & 15019.7 & 22993.7 & 79713.3 &  148544.2 \\
Speed-up & 87.8x  & 142.7x  & 205.6x  &  504.6x &  679.1x   \\
\hline
\end{tabular}
\label{table:Ndet_timings_Hchain}
\end{threeparttable}
\end{table*}

\begin{figure}[h!]
\centering
\includegraphics[width=0.5\textwidth]{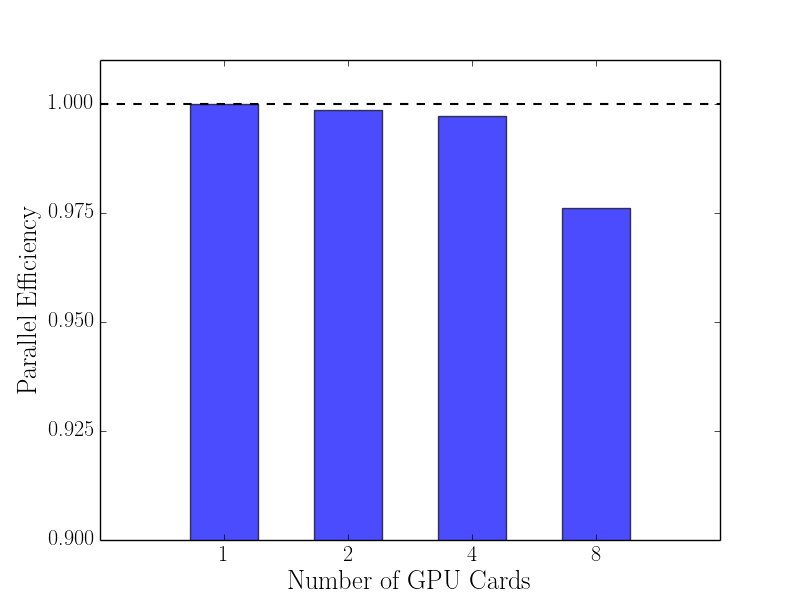}
\caption{Parallel efficiency of our ph-AFQMC code illustrated on H$_{50}$.  We use a CASSCF trial wavefunction with 44 determinants, and 800 walkers propagated for 0.5 $Ha^{-1}$ with $\Delta\tau = 0.01 \ Ha^{-1}$.}
\label{fig:GPU_PE}
\end{figure}

To parallelize across GPU cards on a single node, we divide the total number of walkers into subsets which are independently propagated and measured on different GPU cards.  We use Open Multi-Processing (OpenMP) to achieve shared-memory parallelization of the CPU threads, and to each CPU thread we associate a partner GPU device.  Fig. \ref{fig:GPU_PE} highlights the near-unity parallel efficiency of our implementation, defined as the multi-GPU speed-up over 1 GPU divided by the number of GPUs utilized.   

To treat larger system sizes we have implemented a local memory strategy which spreads slices of the 4-dimensional tensors in Eq. \eqref{Ys} across 8 cards for the entire simulation.  At the intervals where the energy is measured, the random walkers propagated on, e.g., GPU 0 are sent to GPUs 1-7 to compute the components of the two-electron energy derived from the elements stored locally on GPUs 1-7.  This is done simultaneously for walkers on all GPUs, after which the energy components are gathered and tallied for each walker.  The nodes utilized have 8 GPU cards each with 8 GB of RAM.  Using DF and sp, this local memory-slicing algorithm enables us to treat systems as large as $H_{100}$ in the double-zeta basis ($M = $ 500, $N = $ 100).

%----------------------%
\subsection{Accuracy}
%\subsection{IV.C Accuracy}
%----------------------%

To benchmark the accuracy of our algorithm when sp and DF are used, we compute the total energy of $H_{50}$ in the cc-pVDZ basis with a bond length of 1.8 Bohr, and compare with results from a recent study\cite{motta2017Hchain} presenting data from state-of-the-art methods including ph-AFQMC, DMRG, and restricted CCSD(T), among others.  DMRG is essentially exact for one-dimensional systems,\cite{hachmann2006multireference} and RCCSD(T) is expected to provide a high level of accuracy as the bond-length is near its equilibrium value.\cite{purvis1982full} 

Our GPU results are shown in Table \ref{table:Hchain_Simons} along with the previously published data.  The differences in the total energies of sp vs dp for our GPU calculations are 1.2(9) m$Ha$ for DF and 0.7(8) m$Ha$ for CD.  Importantly, both of these are smaller than the resolution required for chemical accuracy (1.6 m$Ha$), confirming that for this system size we can take advantage of the hardware-optimized sp arithmetic on the GPU without incurring a significant loss of accuracy.  
\begin{table}[h]
%\begin{table*}[h!]
%\centering
%\begin{threeparttable}
\caption{Total electronic energies [Ha] of H$_{50}$ at $R=1.8$ Bohr in the cc-pVDZ basis.  Propagation times [hours] are presented using 8 CPU/GPU pairs.  We use 1000 walkers propagated for a length of 200 $Ha^{-1}$ (including equilibration).}  %The CPU dp calculation from Ref. \citenum{motta2017Hchain} was run on AMD Shanghai Opteron processors.
\begin{tabular}{l l l l}
\hline\hline
                        & Electronic Energy &  Propagation Time \\ [0.5ex] 
\hline
GPU sp DF                                 & -125.2107(7) & 14.6 \\ 
GPU dp DF                                 & -125.2119(6) & 31.4 \\
GPU sp Chol                               & -125.2239(6) & 27.8 \\
GPU dp Chol                               & -125.2246(5) & 45.2 \\
CPU dp Ref.~\citenum{motta2017Hchain}     & -125.2242(8) & \\ %$\sim$650  \\
RCCSD(T) Ref.~\citenum{motta2017Hchain}& -125.2067    &      \\
DMRG Ref.~\citenum{motta2017Hchain}    & -125.2210(1) &      \\
\hline
\end{tabular}
\label{table:Hchain_Simons}
%\end{threeparttable}
\end{table}

DF produces about half the number of auxiliary-fields compared with CD (550 vs 1105), reducing propagation times by a factor of 1.9 for sp and 1.4 for dp.  In terms of the resulting accuracy, it is well known that while the DF decomposition may not be sufficient to produce total energies within chemical accuracy, it can recover sub-m$Ha$ accuracy in the calculation of relative energies.\cite{krisiloff2015density,aquilante2007unbiased}  Indeed, we find that for the total energy of H$_{50}$ DF differs from CD with a $10^{-5}$ cutoff by 13.2(9) m$Ha$ in sp and 12.7(8) m$Ha$ in dp, respectively.  Yet to put these errors into context we note in passing that DF ph-AFQMC in both sp and dp produces total energies for this system that are closer to the DMRG reference by $\sim 4$ m$Ha$ than RCCSD(T), known to many as the ``gold standard'' of quantum chemistry.\cite{ramabhadran2013extrapolation,pople1999nobel} 

To conclude this section we illustrate the capacity of DF and more aggressive CD truncation thresholds to recover chemically accurate energy differences for the deprotonation of methanol.  Table \ref{table:MeOH_Habstr_DF} shows errors of $\sim$3 m$Ha$ for the total energies of the neutral and deprotonated species; however the deviation of the energy difference from that of the most stringent CD cutoff is negligible, taking statistical errors into account.  In this molecular case, compared to H$_{50}$, we find a more pronounced reduction in the number of auxiliary-fields, implying a $\sim$4x speed-up (vs $\sim$2x for the hydrogen chain) over CD with a 10$^{-5}$ threshold.  

%\begin{table*}[ht]
%%\begin{table*}[h!]
%%\centering
%\begin{threeparttable}
%\caption{Accuracy of DF and CDs with various cutoffs for the deprotonation energy of methanol.  Sp is used, and long imaginary-time trajectories are stabilized with PC. $N_{AFs}$ denotes the resulting number of auxiliary-fields.}
%\begin{tabular}{l l l l l l}
%\hline\hline
%        & $N_{AFs}$ & MeOH & MeO$^-$ & $\Delta E$ & $\Delta E - \Delta E_{CD 10^{-6}}$ \\ [0.5ex] 
%\hline
%DF           & 142 &  -155.8503(4) & -148.5340(7) & 0.6777(8) & -0.0004(10) \\ 
%CD $10^{-2}$ & 187 &  -155.8144(6) & -148.4985(4) & 0.6772(7) & -0.0008(10) \\
%CD $10^{-3}$ & 385 &  -155.8531(4) & -148.5357(4) & 0.6787(6) & 0.0007(9)   \\
%CD $10^{-4}$ & 471 &  -155.8542(4) & -148.5366(4) & 0.6790(6) & 0.0009(9)   \\
%CD $10^{-5}$ & 617 &  -155.8544(4) & -148.5371(3) & 0.6787(6) & 0.0007(8)   \\
%CD $10^{-6}$ & 855 &  -155.8533(5) & -148.5366(4) & 0.6780(6) & 0           \\
%\hline
%\end{tabular}
%\label{table:MeOH_Habstr_DF}
%\end{threeparttable}
%\end{table*}

\begin{table}[h]
%\begin{table*}[h!]
%\centering
%\begin{threeparttable}
\caption{Accuracy of DF and CDs with various cutoffs for the deprotonation energy of methanol.  Sp is used, and long imaginary-time trajectories are stabilized with PC. $N_{AFs}$ denotes the resulting number of auxiliary-fields.}
\begin{tabular}{l l l l l}
\hline\hline
        & $N_{AFs}$ & MeOH & MeO$^-$ & $\Delta E - \Delta E_{CD 10^{-6}}$ \\ [0.5ex] 
\hline
DF           & 142 &  -155.8503(4) & -148.5340(7) &  -0.0004(10) \\ 
CD $10^{-2}$ & 187 &  -155.8144(6) & -148.4985(4) &  -0.0008(10) \\
CD $10^{-3}$ & 385 &  -155.8531(4) & -148.5357(4) &  0.0007(9)   \\
CD $10^{-4}$ & 471 &  -155.8542(4) & -148.5366(4) &  0.0009(9)   \\
CD $10^{-5}$ & 617 &  -155.8544(4) & -148.5371(3) &  0.0007(8)   \\
CD $10^{-6}$ & 855 &  -155.8533(5) & -148.5366(4) &  0           \\
\hline
\end{tabular}
\label{table:MeOH_Habstr_DF}
%\end{threeparttable}
\end{table}

%%%%%%%%%%%%%%%%%%%%%%%%%%%%%%%%%%%%%%%%%%%
\section{IPs of Transition Metal Atoms}
%\section{V. IPs of Transition Metal Atoms}
%%%%%%%%%%%%%%%%%%%%%%%%%%%%%%%%%%%%%%%%%%%

In this section we compute the IPs of the first-row TM atoms correlating \emph{all} electrons, and compare the calculated ph-AFQMC results to experiment and previous electronic structure calculations.  

%--------------------------------%
\subsection{Computational Details}
%\subsection{V.A Computational Details}
%--------------------------------%

Our computational protocol begins with a restricted (open-shell) HF calculation.  We visually inspect the occupied orbitals of this solution to ensure that the electron configurations shown in Table \ref{table:electron_configs} are obtained.  For some atomic species, HF provides a qualitatively incorrect description of the single-particle orbital occupancies, requiring initialization from custom density matrices to converge subsequent HF calculations to the target ground-state configurations.  We note that for the V$^+$ cation the initial density matrix guess was constructed with the $L=2$ orbital unoccupied.  In all cases the canonical HF orbitals are used to initialize a restricted CASSCF calculation. 

\begin{table*}[ht]
%\centering
\begin{threeparttable}
\caption{Target electron configurations and spin-multiplicities (2S + 1), from Refs. \citenum{thomas2015accurate} and \citenum{balabanov2005systematically}.}
\begin{tabular}{l l l l l l l l l l l}
\hline\hline
  System &   Sc   &   Ti   &   V   &   Cr   &   Mn   &   Fe   &   Co   &   Ni   &   Cu   &   Zn          \\ [0.5ex] 
\hline
Neutral    & 4$s^2$3$d^1$ & 4$s^2$3$d^2$ & 4$s^2$3$d^3$ & 4$s^1$3$d^5$ & 4$s^2$3$d^5$ & 4$s^2$3$d^6$ & 4$s^2$3$d^7$ & 4$s^2$3$d^8$ & 4$s^1$3$d^{10}$ & 4$s^2$3$d^{10}$ \\
Spin Mult. & 2 & 3 & 4 & 7 & 6 & 5 & 4 & 3 & 2 & 1 \\
Cation   & 4$s^1$3$d^1$ & 4$s^1$3$d^2$ & 3$d^4$ & 3$d^5$ & 4$s^1$3$d^5$ & 4$s^1$3$d^6$ & 3$d^8$ & 3$d^9$ & 3$d^{10}$ & 4$s^1$3$d^{10}$ \\
Spin Mult. & 3 & 4 & 5 & 6 & 7 & 6 & 3 & 2 & 1 & 2 \\
\hline
\end{tabular}
\label{table:electron_configs}
\end{threeparttable}
\end{table*}

All ph-AFQMC calculations in this section use a CD cutoff of $10^{-4}$.  We utilize basis sets that have been optimized to account for scalar relativistic effects\cite{balabanov2005systematically}, and use the spin-free exact two-component approach\cite{kutzelnigg2005quasirelativistic,peng2012exact} to decouple the electronic degrees of freedom from the Dirac equation.  This approximation produces one-body terms which we simply add to the non-relativistic Hamiltonian in Eq. \eqref{Hamiltonian}.  

To compare calculations in finite basis sets to experiments we extrapolate the correlation energies to the Complete Basis Set (CBS) limit using two data points fit to $1/x^3$ ($x=3,4$ for TZ,QZ).\cite{purwanto2013frozen,helgaker1997basis,balabanov2005systematically}  We confirmed for a subset of the atoms that the inclusion of the aug-cc-pwCV5Z-DK energies did not significantly change the extrapolated results, consistent with Ref. \citenum{balabanov2005systematically}.  Following Ref.~\citenum{thomas2015accurate} and our own observation that the HF energies converge relatively quickly in this sequence of basis sets, we use the 5Z value for the CBS HF energies. 

The Trotter error due to finite imaginary-time discretization can be extrapolated to 0 using progressively smaller time steps.  Here we use $\Delta\tau = 0.005, 0.01,$ and $0.02 \ Ha^{-1}$.  For Co through Zn we compared the CBS estimate from such an extrapolation with values from the smallest time-step only, $\Delta\tau = 0.005 \ Ha^{-1}$.  In the latter approach we observe a substantial yet systematic cancellation of error, and CBS estimates of equivalent accuracy compared to the 3-point extrapolation approach are shown in Table \ref{table:IP_dt}.  In light of this data we use only $\Delta\tau = 0.005 \ Ha^{-1}$ for all calculations.   

\begin{table}[h!]
\centering
\begin{threeparttable}
\caption{Comparison of CBS IPs [eV] for Co, Ni, Cu, and Zn with $\Delta\tau \rightarrow 0$ vs $\Delta\tau = 0.005 \ Ha^{-1}$ computed in sp with ph-AFQMC/PC.}
\begin{tabular}{l l l l l }
\hline\hline
        &    Co    &    Ni    &   Cu   &   Zn       \\ [0.5ex] 
\hline
Expt.   &   7.87   &   7.59   &  7.73  &  9.39      \\
$\Delta\tau \rightarrow 0$    &   7.87(3)    &   7.61(3)    &   7.54(3)    &   9.33(4)    \\
$\Delta\tau = 0.005 Ha^{-1}$  &   7.89(3)    &   7.59(3)    &   7.55(3)    &   9.37(3)    \\
\hline
\end{tabular}
\label{table:IP_dt}
\end{threeparttable}
\end{table}

Details of the CS procedure can be found in Ref. ~\citenum{shee2017chemical}.  In short, we run a set of independent calculations called repeats, each of which uses a distinct random number seed to propagate both the neutral and cationic species such that pairs of walkers sample the same auxiliary-fields.  After an initial equilibration period, cumulative averages of the energy difference are computed along each of the imaginary-time trajectories, and are averaged among the set of repeats to obtain an estimate of statistical error.  In the present case, stochastic error cancellation leads to a pronounced reduction in the variance of the IPs, and allows converged measurements to be attained at short imaginary times.  We will show in the next section that this leads not only to significant reductions in computational cost, relative to the uncorrelated approach, but also to systematically improved accuracy.

%---------------------------------%
\subsection{Results and Discussion}
%\subsection{V.B Results and Discussion}
%---------------------------------%

Tables \ref{table:Sc_Mn_compare_expt} and \ref{table:Fe_Zn_compare_expt} summarize our results for the all-electron IPs of the first-row TM atoms.  We show values obtained from both PC and CS ph-AFQMC approaches, and compare with experimental and CCSD(T) values.

\begin{table}[h]
%\begin{table*}[h!]
%\centering
%\begin{threeparttable}
\caption{Calculated ph-AFQMC IPs [eV] in the CBS limit computed with sp and $\Delta\tau = 0.005 \ Ha^{-1}$, compared with experimental and CCSD(T) values. Experimental IPs have spin-orbit contributions removed.}
\begin{tabular}{l l l l l l}
\hline\hline
                    &  Sc      &  Ti       &  V         &  Cr        &  Mn           \\ [0.5ex] 
\hline
%CASSCF              & & & & &                           \\
ph-AFQMC/PC         & 6.51(1)  & 6.71(2)   & 6.74(1) & 6.75(2) & 7.41(2)  \\
ph-AFQMC/CS         & 6.52(3)  & 6.80(3)   & 6.74(3) & 6.74(3) & 7.45(3)  \\
Expt.               & 6.56     & 6.83      & 6.73    & 6.77    & 7.43     \\
CCSD(T)\tnote{*}    & 6.54     & 6.81      & 6.73    & 6.79    & 7.42     \\    
\hline
\end{tabular}
\begin{tablenotes}
\footnotesize 
\item[*] Ref. \citenum{balabanov2005systematically}
\end{tablenotes}
\label{table:Sc_Mn_compare_expt}
%\end{threeparttable}
\end{table}

\begin{table}[h]
%\begin{table*}[h!]
%\centering
%\begin{threeparttable}
\caption{Same as Table \ref{table:Sc_Mn_compare_expt}, but for atoms in the right-half of the row.}
\begin{tabular}{l l l l l l}
\hline\hline
                    &  Fe  &  Co  &  Ni  &  Cu  &  Zn    \\ [0.5ex] 
\hline
%CASSCF             &  6.87      &  5.30      &  6.27        & &           \\
ph-AFQMC/PC         &  7.86(2)   &  7.89(3)   &  7.59(3)     & 7.55(3)    &  9.37(3)      \\
ph-AFQMC/CS   &  7.89(2)   &  7.87(3)   &  7.61(2)     & 7.68(3)    &  9.37(3)      \\
Expt.               &  7.90      &  7.87      &  7.59        & 7.73       &  9.39         \\
CCSD(T)\tnote{*}    &  7.89      &  7.88      &  7.59        & 7.72       &  9.37         \\    
\hline
\end{tabular}
\label{table:Fe_Zn_compare_expt}
%\end{threeparttable}
\end{table}

\begin{table*}[ht]
%\begin{table*}[h!]
%\centering
\begin{threeparttable}
\caption{Number of active electrons and orbitals in the CASSCF trial wavefunctions for the cation/neutral species, and the number of determinants kept in the ph-AFQMC trial function accounting for 99.5$\%$ of the CI weight. For all species in this table the 3p electrons are active.}
\begin{tabular}{l l l l l l}
\hline\hline
                    &  Sc  &  Ti  &  V  &  Cr  &  Mn \\ [0.5ex] 
\hline
Active Space & 8/9e,16o & 9/10e,16o & 10/11e,19o & 11/12e,16o & 12/13e,18o\tnote{\dag} \\
%CASSCF             &  6.87      &  5.30      &  6.27        & &           \\
$N_{dets}$ TZ &  146/224 & 240/442 & 366/751 & 303/271 &  423/584 \\ 
$N_{dets}$ QZ &  143/439 & 293/388 & 300/903 & 92/262  &  852/1266 \\ 
\hline
\end{tabular}
\begin{tablenotes}
\footnotesize 
\item[\dag]  Three 5$p$ orbitals replaced by five 4$d$ orbitals in the active space.
\end{tablenotes}
\label{table:ActiveSpacesLeftHalf}
\end{threeparttable}
\end{table*}

\begin{table*}[ht]
%\begin{table*}[h!]
%\centering
\begin{threeparttable}
\caption{Same as Table \ref{table:ActiveSpacesLeftHalf}, but for Cu and Zn with 99.0$\%$ of the CI weight retained.}
\begin{tabular}{l l l l l l}
\hline\hline
                    &  Fe  &  Co  &  Ni  &  Cu  &  Zn    \\ [0.5ex] 
\hline
Active Space &  7/8e,18o  &  8/9e,13o  &  9/10e,13o   & 10/11e,18o & 11/12e,13o \\
%CASSCF             &  6.87      &  5.30      &  6.27        & &           \\
$N_{dets}$ TZ &  23/227 & 210/85 &  138/156 & 374/322 &  299/518 \\ 
$N_{dets}$ QZ &  23/121 & 237/66 &  159/161 & 504/507 &  277/526 \\ 
\hline
\end{tabular}
\label{table:ActiveSpacesRightHalf}
\end{threeparttable}
\end{table*}

The active spaces employed for the neutral and cationic species in the TZ and QZ basis sets are described in detail in Tables \ref{table:ActiveSpacesLeftHalf} and \ref{table:ActiveSpacesRightHalf}.  In general, the use of truncated CASSCF trial wavefunctions in ph-AFQMC involves subtleties that require careful consideration, since the CASSCF calculation itself becomes a pre-processing element of higher computational scaling, and the truncation breaks size extensivity.  This approach is viable if the ph-AFQMC result converges quickly with trial wavefunctions generated from active spaces much smaller than the full Hilbert space.  For atoms and molecules this is typically the case, and an internal validation procedure within ph-AFQMC can be employed involving a series of calculations using various active space and trunctation cutoffs.  In particular, for Fe-Zn we started by including the 4$s$ and 3$d$ electrons in an active space composed of 13 active orbitals.  While the resulting truncated CASSCF trial wavefunctions produced sufficiently accurate ph-AFQMC/PC results in the CBS and $\Delta\tau \rightarrow 0$ limits for Co, Ni, and Zn, 18 orbitals were required in the case of Fe.  The improvement in the IP resulting from the inclusion of a second shell of $d$ orbitals in the CASSCF active space is a manifestation of the so-called ``double-shell'' effect.\cite{bauschlicher1988atomic,andersson1992excitation}  We find that this effect is less pronounced in the case of all-electron ph-AFQMC since the application of $e^{-\tau \hat{H}}$ to walker configurations can explore the space of excitations into virtual $d$ orbitals even if such excitations are not represented in the trial function.       

For the left half of the 1st row of transition metals in the periodic table, Sc-Mn, we designate the 3$p$ electrons as active in addition to the 4$s$ and 3$d$ electrons.  In an effort to maintain consistency (i.e. to include HF virtuals of similar character in the initial guesses for the CASSCF procedure) among all atoms in the row, for those in the left-half we start with 16 active orbitals.  This produced accurate ph-AFQMC/PC results for Sc and Cr.  For V we noticed a sharp drop in energy in both the neutral and cationic species going from 16 to 19 active orbitals; for Mn an accurate IP required the replacement of three 5$p$ orbitals with five 4$d$ in the CASSCF active space to accommodate the double-shell effect.   

The case of Cu proves to be particularly challenging, and illustrates an additional merit of the CS approach.  A trial function with 18 active orbitals approaches the memory limit of traditional CASSCF solvers, but is still insufficient to produce results of the desired accuracy within ph-AFQMC/PC.  With additional active orbitals, approximate CASSCF solvers utilizing DMRG\cite{sun2017general} did converge, but only a subset of the resulting configurations and CI coefficients could be accessed with the current implementation of selected CI in PySCF.  Even with moderate selection cutoffs, when such a wavefunction was used as a trial function in ph-AFQMC we found a significant increase in statistical error, in addition to larger deviations of the resulting IP from experiment.  

\begin{figure}[ht]
%\begin{figure}[h!]
\centering
\includegraphics[width=0.5\textwidth]{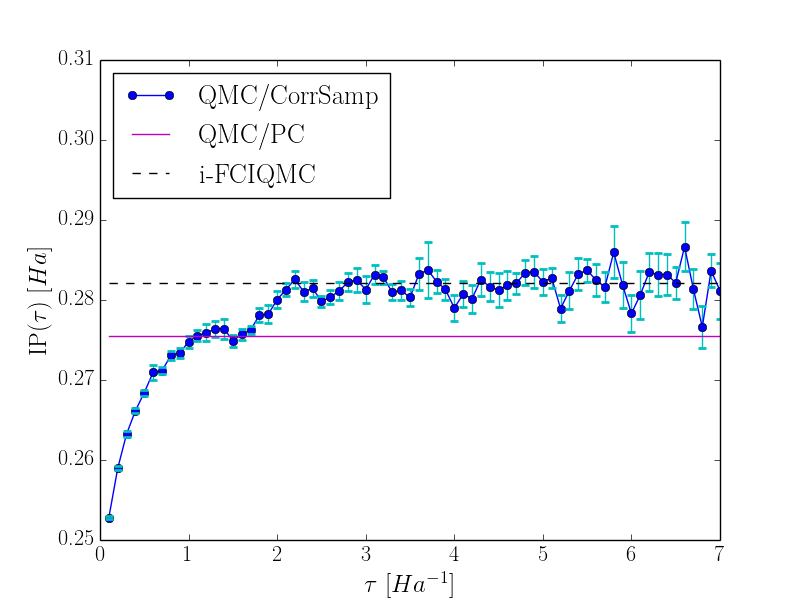}
\caption{Comparison of the IP of Cu as a function of imaginary-time produced from ph-AFQMC/CS in the aug-cc-pwCVQZ-DK basis compared to the regular ph-AFQMC/PC result in the same basis.  The i-FCIQMC result in the aug-cc-pVQZ-DK basis is indicated by a dashed line.}
\label{fig:Cu_FCIQMC}
\end{figure}

In contrast to regular ph-AFQMC/PC, which stabilizes long imaginary-time trajectories, a key advantage of the CS approach is that averaging among independent repeats at short times allows for not only a vast variance reduction when the auxiliary fields are correlated, but also the ability to converge measurements of the energy difference \emph{before} the full onset of the bias that results from the phaseless constraint.  Even though the phaseless approximation is made after each time step, the walker weights at early times stay relatively closer to their true unconstrained values than at long times when the phaseless constraint has fully equilibrated.  To illustrate this we plot the IP of Cu in the TZ basis at short imaginary-times in Fig. \ref{fig:Cu_FCIQMC}.  At longer imaginary times (not shown), the CS IP eventually returns to the ph-AFQMC/PC result, yet from 2-7 $Ha^{-1}$, ph-AFQMC/CS unambiguously converges on an answer consistent with iFCI-QMC, which is expected to be very accurate here,\cite{thomas2015accurate} and moreover after CBS extrapolation is within range of chemical accuracy with respect to experiment.  

The case of Ni is also quite remarkable.  Both CS and PC methods produce IPs consistent with the experimental value and each other in the CBS limit, however a detailed comparison with CCSD(T) values in each basis set, shown in Table \ref{table:Ni_basis}, reveals that this agreement is due to fortuitous cancellations of error.  While the CCSD(T) values approach the CBS limit from above, the ph-AFQMC/PC values approach the same value from below.  ph-AFQMC/CS calculations, on the other hand, produce statistically consistent results with CCSD(T) in each basis and in the CBS limit.

\begin{table}[h]
%\begin{table*}[h!]
%\centering
%\begin{threeparttable}
\caption{Comparison of ph-AFQMC IPs for Ni, as obtained with CS and regular ph-AFQMC/PC, with CCSD(T) in triple- and quadruple- zeta basis sets and in the CBS limit.\cite{balabanov2005systematically}  The CCSD(T) values were obtained with a composite method, namely cc-pVxZ-DK results plus a core-valence correction, which is the difference in the cc-pwCVxZ-DK basis of CCSD(T) calculations with active spaces defined by 3$s$3$p$3$d$4$s$ and 3$d$4$s$ orbitals (x=T,Q).  QMC results used the aug-cc-pwCVxZ-DK basis sets.}
\begin{tabular}{l l l l l l}
\hline\hline
   & CCSD(T) &  $\Delta$QMC/CS & $\Delta$QMC/PC  \\ [0.5ex] 
\hline
TZ  &  7.68 & 7.68(1) & 7.56(2)  \\ 
QZ  &  7.63 & 7.64(1) & 7.57(1)  \\ 
CBS &  7.59 & 7.61(2) & 7.59(3)  \\
\hline
\end{tabular}
\label{table:Ni_basis}
%\end{threeparttable}
\end{table}
%\end{table*}

For the case of Ti we chose to use CS as an alternative to increasing the size of the active space.  Generally, for all atoms in this work CS results exhibit equivalent or better accuracy compared to the conventional method of running ph-AFQMC calculations with PC.  Moreover, the ability to consistently produce chemically accurate results while using sp is reassuring, given that the total energies involved in these calculations are on the order of $\sim$ -1000 $Ha$.  This implies that m$Ha$ energy scales require precision out to at least 7 significant figures, which would be stretching the typical capabilities of sp arithmetic in deterministic algorithms. 

Table \ref{table:TM_timings} shows the total propagation-times required to produce the final IPs in Tables \ref{table:Sc_Mn_compare_expt} and \ref{table:Fe_Zn_compare_expt}, which account for calculations of the total energies of the neutral and cation in the TZ and QZ basis sets.   8-core CPU times are estimated by scaling the propagation time of a small 20 walker system propagated for 1 $Ha^{-1}$ by the required factors to reproduce the parameters of the GPU/PC calculations, i.e. 2000 walkers propagated in the TZ/QZ bases for 120/130 $Ha^{-1}$ for Cr and Fe, and for 200/230 $Ha^{-1}$ for Cu.  We assume perfect parallel efficiency in projecting our single-core CPU estimates to 8-cores.  Obtaining comparable error bars using CS with our GPU code required the propagation of 200 walkers in the TZ/QZ bases for 5/3 $Ha^{-1}$ for Cr, 4/3 $Ha^{-1}$ for Cr, and 10/6 $Ha^{-1}$ for Cu.  For Cr and Cu we use 16 repeats in both basis sets, and for Fe we found that only 5/8 repeats in TZ/QZ are needed.  We note that the larger wall-times for Cu are due to 1) the larger number of both particles and determinants in the trial functions employed, and 2) the relatively poor trial function (compared to the exact ground state) which leads to longer propagation and equilibration times in the PC and CS methods, respectively.  It may be the case that the relatively small atomic radius of Cu results in larger dynamical correlations compared to the rest of the atoms in the row, which are unaccounted for in the CASSCF trial wavefunctions (this explanation is consistent with the relative difficulty we encountered previously in calculating the electron affinity of the flourine atom \cite{shee2017chemical}).  The corresponding speed-ups for these selected atoms are shown in Table \ref{table:TM_speedups}.  

\begin{table}[h]
%\begin{table*}[h!]
%\centering
%\begin{threeparttable}
\caption{Total propagation times [hours] required to produce the final all-electron ph-AFQMC IPs in this work.}
\begin{tabular}{l l l l}
\hline\hline
               &    Cr          &   Fe     &   Cu    \\ [0.5ex] 
\hline
Est. 8-core CPU PC dp   & 9784.5 & 6966.9 & 34150.3  \\
8-card GPU PC sp        & 50.5   & 47.5   & 136.4    \\
8-card GPU CS sp        & 2.4    & 0.9    & 6.0      \\
\hline
\end{tabular}
\label{table:TM_timings}
%\end{threeparttable}
\end{table}
%\end{table*}

\begin{table}[ht]
%\begin{table*}[h!]
%\centering
%\begin{threeparttable}
\caption{Speed-ups corresponding to the timings in Table \ref{table:TM_timings}.  All CPU/GPU calculations use dp/sp respectively.}
\begin{tabular}{l l l l}
\hline\hline
                          &  Cr   &   Fe  &   Cu  \\ [0.5ex] 
\hline
GPU PC vs CPU PC  & 194x  & 138x  & 250x  \\
GPU CS vs GPU PC  & 21x   & 53x   & 23x   \\
GPU CS vs CPU PC  & 4100x & 7700x & 5700x \\
%GPU CS over CPU PC  & 4077x & 7741x & 5692x \\
\hline
\end{tabular}
\label{table:TM_speedups}
%\end{threeparttable}
\end{table}
%\end{table*}

%----------------%
% Final Comments %
%----------------%
We conclude this section with a few remarks.  Currently we use a simple combing method\cite{nguyen2014cpmc} to implement PC.  More sophisticated schemes are possible which may improve the statistical accuracy of the calculation.  While this would slightly reduce the wall-times for the ph-AFQMC/PC method, the accuracy of the results with respect to experiments will be unchanged, since any bias due to PC vanishes when a large population ($\sim$2000 walkers) is used.  At the time of writing, auxiliary basis sets optimized for the scalar relativistic Hamiltonian and DK basis sets used in this work are not publicly available.  The ability to use a  DF decomposition would certainly provide additional speed-ups, although its effect on accuracy remains to be tested for these TM systems.  Finally, we note that the capacity of our GPU code to treat $O$(1000) determinants in the trial wavefunction will likely enable the accurate study of many strongly-correlated systems.  We anticipate that fewer determinants will be needed for metal-ligand complexes (in which the ligand is a non-metal), as TM atoms typically exhibit larger static correlation effects than most coordinated complexes.  In addition, the use of symmetry constraints in the CASSCF calculations will greatly reduce the number of configurations in the CI expansions.

%%%%%%%%%%%%%%%%%%%%
\section{Conclusions and Outlook}
%\section{VI. Conclusions and Outlook}
%%%%%%%%%%%%%%%%%%%%

We have designed a GPU implementation of ph-AFQMC for single- and multi-determinant trial wavefunctions which can drastically reduce the scaling prefactor in realistic electronic structure calculations with near-unity parallel efficiency.  Our strategy utilizes new batched SMW and energy algorithms, along with the use of sp and the DF decomposition.  We validate performance enhancement with ph-AFQMC calculations of linear chains of hydrogen atoms and the atomic IPs of Sc through Zn, finding speed-ups relative to the CPU in dp of two orders of magnitude and which increase with the number of determinants in the trial wavefunction. For H$_{50}$ and TM IPs, sp is sufficient to produce accuracy on the scale of 1 kcal/mol with respect to exact methods and experiment, respectively.  In this work we also demonstrate that our previously outlined CS approach to ph-AFQMC enables both additional speed-ups of an order of magnitude, as well as the ability to converge measurements before the full onset of the bias due to the phaseless constraint.  For all TM atoms, CS produces equivalent and often more accurate IPs, and in a fraction of the wall-time.  

We are optimistic that in the near future the investigation of many large, realistic systems will be feasible with ph-AFQMC.  In what follows we anticipate issues of scalability and describe the possible solutions we envision.  To enable large-scale calculations that efficiently utilize available High-Performance Computing clusters, we have designed a simple and scalable scheme to parallelize \emph{across} GPU nodes. Once a small walker population is equilibrated, data can be copied to all available nodes and used to initialize independent trajectories on separate nodes, each with a different random number seed.  These sub-trajectories can later be combined into a single trajectory from which averages and error bars can be obtained.

We note that our current memory limitation is rather artificial in the sense that GPU architectures and computing capabilities are improving at a rapid pace, suggesting that the memory capacity of GPU cards will continue to increase.  Also, Message Passing Interface (MPI) could be used to extend our local memory-slice scheme to multiple nodes.  Finally, we note the plausibility of utilizing CPU memory to store high-dimensional tensors, relying on rapidly improving data transfer rates to send slices to the GPUs.  At the time of writing, Nvidia's NVLink boasts speeds of $\sim$300 GB/sec on Tesla V100 GPUs.  This technology provides a shared virtual memory space and in time may completely eliminate the need to explicitly transfer information from host memory to GPU memory or between two GPU cards on the same node.

The incorporation of additional theoretical approaches that would capture the same level of accuracy with diminished computational cost are also currently under consideration.  One possibility is the use of Canonical Transcorrelation theory to produce an effective Hamiltonian with the electron cusps analytically removed,\cite{yanai2012canonical} which is related to the Jastrow factor in DMC.  We anticipate that this will reduce the number basis functions required to reach the CBS limit.\cite{sharma2014spectroscopic,knizia2009simplified}  In addition, one may exploit real-space locality in the context of electronic excitations, which also has the potential to drastically reduce both the current scaling and memory demands.\cite{murphy1995pseudospectral,saitow2017new,riplinger2013efficient}  Lastly, we note that when sp is insufficient, mixed-precision matrix strategies, which are well-studied and relatively straightforward to implement,\cite{goddeke2007performance,olivares2009accelerating,esler2012accelerating} may be employed.

Combining the speed-ups due to the GPU and CS, we now have a robust and efficient computational protocol that is approximately three orders of magnitude faster than previous AFQMC procedures for large systems.  Once the memory bottleneck is alleviated with the strategies mentioned above, systems that previously were inaccessible to study with ph-AFQMC will soon be within reach.  Future targets include low-energy redox- and spin- states of catalytic metalloporphyrins\cite{bykov2015six,shen2016dft} and iron-sulfur clusters,\cite{sharma2014low} and the computation of $pKa$s for the oxygen evolving complex of Photosystem II.\cite{askerka2016o2}  
%We also intend to benchmark redox potentials and spin-splitting energies of a large set of TM complexes, essentially creating \emph{ab initio} data of quality approaching that of experiments that might be used to train deep learning methods or e.g. DFT correction schemes.\cite{friesner2017localized}  Finally, we intend to compute the charge gap\cite{vitali2016computation} of a copper oxide sheet via a fully atomistic simulation with periodic boundary conditions. 
Our method may also open the door to the accurate and fully \emph{ab initio} investigation of strongly correlated solids such as high-temperature superconducting materials.\cite{orenstein2000advances,lee2006doping}  These and other targets of study will be the subject of future investigations. 

\section{Acknowledgements}

% ACS style
%\begin{acknowledgement}

JS gratefully acknowledges Mario Motta and Hao Shi for providing scripts to initialize HF calculations of hydrogen chains and V$^+$, respectively, and Qiming Sun for help with PySCF.  JS would also like to thank Roald Hoffmann  and Kirk A. Peterson for insightful discussions about TM atoms.  DRR acknowledges funding from NSF CHE-1464802; SZ from NSF DMR-1409510.

% ACS style
%\end{acknowledgement}

%\begin{suppinfo}

%This will usually read something like: ``Experimental procedures and
%characterization data for all new compounds. The class will
%automatically add a sentence pointing to the information on-line:

%\end{suppinfo}

%\bibliography{achemso-demo}
\bibliography{References}

\end{document}